\documentclass[12pt]{article}
\usepackage{graphicx}
\usepackage{epsfig}
\usepackage{amsmath}
\usepackage{amsthm}
\usepackage[round]{natbib}

\newcommand{\n}[1] {\mbox{\boldmath{$#1$}}}
\newcommand{\be}{\begin{eqnarray}}
\newcommand{\ee}{\end{eqnarray}}
\newcommand{\beq}[1]{\begin{equation}\label{#1}}
\newcommand{\eeq}{\end{equation}}
\newcommand{\ba}{\begin{eqnarray*}}
\newcommand{\ea}{\end{eqnarray*}}
\DeclareMathOperator*{\argmax}{arg\,max}

\title{Inferences in Bayesian variable selection problems with large model spaces}

\author{Garc\'{\i}a-Donato, G$^1$ and Mart\'{i}nez-Beneito, MA$^2$\\
\footnotesize{$^1$  Universidad de Castilla La Mancha, Spain,  $^2$ CSISP Valencia, Spain}}
\date{\today}
\begin{document}
\maketitle

\graphicspath{{plots//}}

\begin{abstract}
An important aspect of Bayesian model selection is how to deal with huge model spaces, since exhaustive enumeration of all the models entertained is unfeasible and inferences have to be based on the very small proportion of models visited. This is the case for the variable selection problem, with a moderate to large number of possible explanatory variables being considered in this paper. We review some of the strategies proposed in the literature and argue that inferences based on empirical frequencies via Markov Chain Monte Carlo sampling of the posterior distribution outperforms recently proposed searching methods. We give a plausible yet very simple explanation of this effect, showing that estimators based on frequencies are unbiased. The results obtained in two illustrative examples provide strong evidence in favor of our arguments.\\

\end{abstract}

\section{Inferences in large model spaces}\label{sec.est}
This paper is rooted in the model selection problem, that is with uncertainty surrounding the probabilistic model which, from an initial set ${\cal M}$ of candidates, better explains certain data $\n y$. In particular, we address the variable selection problem where the competing models differ about which subset of variables are to be included as explanatory covariates for $\n y$. 

One special characteristic of the variable selection problem is that  ${\cal M}$, the model space, easily becomes extremely large. For instance, a problem with $p=40$ potential covariates has $2^{40}\approx 10^{12}$ different models. The mere binary representation of such a model space would occupy 5 terabytes of memory.

We focus on the difficulties that arise as a consequence of the very large size of  ${\cal M}$. We consider the problem from a Bayesian point of view and the context we use for the development of our ideas is the problem of variable selection in Gaussian regression models.

The Bayesian approach to the problem is conceptually straightforward. Any feature of interest, say $\tau$, is a deterministic function of the posterior distribution over the model space. Examples of such features are the highest posterior probability model (hereafter HPM), the inclusion probabilities of covariates or posterior predictions of a new value of the dependent variable. Unfortunately, three major difficulties arise when putting the Bayesian approach into practice: i) the choice of the prior distributions; ii) the computation of the integrated likelihood (or equivalently the Bayes factors) for single models in ${\cal M}$ and, in large model spaces, iii) the {\em estimation} of $\tau$, since its exact value is virtually unknown due to the size of  ${\cal M}$. Benchmark papers for each of these areas of research are respectively, \citet{BerPer01}, \cite{ChibJeliazkov01} and \cite{GeorgeMcCulloch97}.

Our work is basically concerned with iii), which is intimately related with strategies for exploring the model space (i.e. visiting a small proportion of models, hereafter denoted ${\cal M}^*$), since covering the whole model space is unfeasible. Our main aim is to shed new light on a topic (almost an implicit debate) that from time to time appears in the literature (see references below). The subject is about the estimation of $\tau$ and more concisely whether it should be based on the {\em empirical} distribution (observed frequencies in ${\cal M}^*$) or on the normalized Bayes factors of models in  ${\cal M}^*$. In the first approach, until quite recently the general one, models in  ${\cal M}^*$ are visited according to a sampling scheme with the posterior distribution of the models in ${\cal M}$ as the stationary distribution. Markov Chain Monte Carlo methods are commonly used for this task. In the second approach the emphasis is placed on visiting, usually without replacement, good models (i.e. with high posterior probability). In the rest of the paper, for ease of comprehension, we respectively refer to {\em empirical} and {\em re-normalized} for each of the approaches outlined above. The common use of {\em empirical} methods is MCMC methods for sampling from the posterior distribution plus an estimation of $\tau$ via frequencies. On the other hand, the {\em re-normalized} approach uses algorithms for sampling good models and estimations which are obtained via the {\em re-normalized} analytical expression of Bayes factors. {\em Empirical} methods have been proposed and used by \cite{GeoMc93}, \cite{GeorgeMcCulloch97}, \cite{KuoMal98}, \cite{Deletal00}, \cite{NottKohn05}, \cite{Nt02}, \cite{Nt09} and \cite{CasMor06} (just to mention a few). Papers more in favor of the {\em re-normalized} approach are \cite{Clyetal10}, \cite{BerMol05}, \cite{CarSc09} and \cite{ScCar09}.

{\em Re-normalized} methods are motivated by the sound argument that the frequency of visits, in such huge model spaces, is a poor basis for estimation since the number of repeated visits (if any) is very small. Several authors (see eg. \citealt{Clyetal10} and \citealt{ScCar09}) have argued in favor of the superiority of these procedures over the {\em empirical} ones. Nevertheless, as we further explain, our experience is quite the opposite, finding that in general {\em empirical} estimations outperform their {\em re-normalized} counterparts in key aspects. Our explanation for this effect is simple: {\em empirical} estimators are particular cases of {\em probability proportional to size sampling (PPS) estimators} (see \citealt{Lohr99}), and hence unbiased. Furthermore, these estimators have an associated measure of precision which can be very useful for the problem at hand. An appealing and well known extra property of {\em empirical} methods is that, although exploring high probability models is not their ultimate goal, these appear more frequently simply because they are more probable.

Throughout this paper we develop and formalize the ideas outlined above. With this aim in mind, the  paper has been organized in the following way.

In Section~\ref{Zellner} we formulate the problem of the variable selection considered. In order to keep the impact of difficulties i) and ii) above under control, we use Zellner's $g$-priors \citep{Zellner86} which produce Bayes factors in a closed-form. The corresponding formulae are also briefly  given in Section~\ref{Zellner}. In Section~\ref{sec.slm} the usual methods for obtaining ${\cal M}^*$ are reviewed. In Section~\ref{pss}, {\em empirical} estimators are expressed as PPS estimators and several properties are shown and discussed. In Section~\ref{sec.exact} we present the exact results of a moderately large problem (Ozone35) with $p=35$ covariates obtained with parallel computing, programmed for the occasion with optimized C routines. We then compare these exact results with those obtained with {\em empirical} and {\em re-normalized} estimators, showing strong evidence in favor of the first ones. In Section~\ref{sec.larger} we analyze a much larger dataset (Ozone65) with 65 covariates (for which we do not have the exact answer). Finally, Section~\ref{sec.Ext} contains a summary of the main conclusions in this paper. 

\section{Bayesian variable selection}\label{Zellner}
Let $\n X=\{x_{ij}\}$ be an $N\times p$ full rank matrix and $\n\gamma=(\gamma_1,\ldots,\gamma_p)$ be a $p$-dimensional vector of binary variables. Denote $k_\gamma=\sum\gamma_i$ and for each $\n\gamma$, let $\n X_\gamma$ be the $N\times k_\gamma$ design matrix corresponding to the columns with ones in $\n\gamma$.

The variable selection problem we consider has $2^p$ competing models, each proposed as a plausible explanation of an $N$ dimensional vector $\n Y$. More concisely
\begin{equation}\label{TheProblem}
M_\gamma:\n Y\sim N_N(\alpha\n 1+\n X_\gamma\n\beta_\gamma,\sigma^2\n I),
\hspace{.2cm}\n\gamma\in\{0,1\}^p.
\end{equation}
In this problem, the model space ${\cal M}$ can be represented by $\{0,1\}^p$. The simplest model among the proposed ones is
$$
M_0:\n Y\sim N_N(\alpha\n 1,\sigma^2\n I).
$$

Without loss of generality, posterior probabilities of the models can be expressed as
\begin{equation}\label{postprob}
Pr(M_\gamma\mid \n y)=C\,B_{\gamma 0}\, Pr(M_\gamma),
\end{equation}
where $B_{\gamma 0}$ is the Bayes factor of $M_\gamma$ to $M_0$, $Pr(M_\gamma)$ is the prior probability of $M_\gamma$ and
$$
C^{-1}=\sum_\gamma\, B_{\gamma 0}\,Pr(M_\gamma),
$$
is the constant of proportionality.

Bayes factors are the ratio of the marginal prior predictive distributions evaluated at $\n y$, that is, $B_{\gamma 0}=m_\gamma(\n y)/m_0(\n y)$, where
$$
m_\gamma(\n y)=\int N_N(\n y\mid \alpha\n 1+\n X_\gamma\n\beta_\gamma,\sigma^2\n I)\, \pi_\gamma(\alpha,\n\beta_\gamma,\sigma)\, d\alpha\, d\n\beta_\gamma\, d\sigma.
$$
The function $\pi_\gamma$ is the prior distribution for the parameters under model $M_\gamma$. It is well known that this prior can be neither improper nor {\em vague} (with a very large variance) since the resulting Bayes factors are essentially arbitrary (see \citealt{BerPer01}). Our default choice for this prior is the $g$-prior proposed by \citet{Zellner86}:
$$
\pi_\gamma(\alpha,\n\beta_\gamma,\sigma \mid g)=\sigma^{-1}\, N_{k_\gamma}(\n\beta_\gamma\mid \n 0,g\sigma^2\, (\n X_\gamma^t(\n I-{N}^{-1}\n 1\n 1^t)\n X_\gamma)^{-1}),
$$
for $\n\gamma\ne\n 0$ and $\pi_0(\alpha,\sigma)=\sigma^{-1}$ for $M_0$.

The $g$-priors seem to be greatly inspired by Jeffreys' ideas \citep{Jef61} and the corresponding extension to regression problems in \cite{ZellSiow80} and \cite{ZellSiow84}. The assignment of the constant $g$ has been analyzed by several authors (see \citealt{liang08} and references therein). This parameter $g$ must increase with $N$ to avoid an asymptotically degenerate prior. The default assignment $g=N$ gives rise to a `unit information" prior in the sense that the covariance matrix is corrected by the sample size (see \citealt{Raf98}). 

An alternative to the choice of the constant $g$ is to assume, hierarchically, a proper prior on $g$, say $\pi(g\mid\n\gamma)$. In general, the resulting prior for $\n\beta_\gamma$ has heavy tails, this being an appealing characteristic of a model selection prior which is related to properties like information consistency (\citealt{BayGar08}). Examples of such priors are the multivariate Cauchy for $\n\beta_\gamma$ proposed by Jeffreys-Zellner-Siow (\citealt{Jef61}, \citealt{ZellSiow80} and \citealt{ZellSiow84}) which corresponds to $\pi(g\mid\n\gamma)=Gamma^{-1}(1/2,N/2)$, the hyper-$g$ priors of \cite{liang08}, the Conventional Robust prior in \cite{forte08} and the extension of the $g$ prior in \cite{MarGeo10}.


The $g$-prior provides closed-form expressions for the Bayes factors. In fact it can easily be shown that
\begin{equation}\label{gBF}
B_{\gamma 0}(g)=\Big(1+g\frac{SSE_\gamma}{SSE_0}\Big)^{-(N-1)/2}\,(1+g)^{(N-k_\gamma-1)/2},
\end{equation}
where $SSE_\gamma$ is the sum of the squared errors for $M_\gamma$. Therefore, if all models $\n\gamma$ can be visited, their posterior probabilities can be calculated without great computational effort. Interestingly, the proposals in \cite{forte08} and \cite{MarGeo10} also lead to closed-form Bayes factors.

Notice that independently of the approach adopted to construct the inferences (either {\em empirical} or renormalized) the above expression can easily be used to unequivocally identify the model that, within a given set of models, has the largest posterior probability (since of course it coincides with the model with the largest $B_{\gamma 0}(g)$).

\section{Search in large model spaces}\label{sec.slm}
With the distinction introduced in Section~\ref{sec.est}, {\em empirical} methods use the relative frequencies of models visited in a  subset ${\cal M}^*\subset{\cal M}$ as the basis for the estimation of $\tau$ (the quantity of interest). On the other hand, {\em re-normalized} methods base this estimation on the use of the renormalized expression of Bayes factors for models in ${\cal M}^*$. Clearly, the way ${\cal M}^*$ is obtained can vary from one approach to the other. In this section we succinctly overview the methods used in the two approaches.

\paragraph{${\cal M}^*$ in {\em empirical} methods}
Markov Chain Monte Carlo (MCMC) methods have provided decisive numerical support for the development of Bayesian methods over the last two decades. Bayesian model selection is not an exception and the literature devoted to MCMC strategies for solving the problem is extensive. When estimations are based on the frequency of visits, the models visited form approximately a sample from the posterior distribution. In this setting MCMC methods are an essential tool for generating the sample  mentioned.

A great majority of these proposals are to a certain extent based on the seminal work by \cite{GeoMc93}, greatly improved and extended in \cite{GeorgeMcCulloch97}. A number of interesting contributions on this area are \cite{KuoMal98}, \cite{Deletal00}, \cite{NottKohn05}, \cite{Nt02}, \cite{Nt09} and \cite{CasMor06}.

In Appendix~\ref{ApGS} we describe the sampling strategy we propose for the model selection problem in (\ref{TheProblem}) with hierarchical $g$ priors. This is a straightforward Gibbs sampling scheme that takes advantage of the integrated expression in (\ref{gBF}). The particular case for $g$-priors that we used in the examples had already been suggested by \cite{GeorgeMcCulloch97}. This sampling scheme can become extremely efficient in combination with updating identities for the SSE's (see \citealt{Gentle07} and references therein) since it is built upon steps in which a variable is either added or deleted.

With an MCMC sampling, given an initial model $\n\gamma^{(0)}$, we obtain a sample of models ${\cal M}^*=\{\n\gamma^{(1)}, \n\gamma^{(2)},\ldots,\n\gamma^{(n)}\}$ having $Pr(M_\gamma\mid\n y)$ as the stationary distribution. This is a key characteristic of ${\cal M}^*$ that provides the ensuing {\em empirical} estimators of $\tau$ with important characteristics described in Section~\ref{pss}.

\paragraph{${\cal M}^*$ in {\em re-normalized} methods}
The origins of this approach date back at least to \cite{GeorgeMcCulloch97} who pointed out the possibility of using an MCMC sample ${\cal M}^*$ in combination with the normalized expression of Bayes factors as the basis for producing the required inferences. For those models in ${\cal M}$ not sampled, the posterior probability is assumed to be zero and for the rest $\widehat{Pr}(M_\gamma\mid\n y)\propto B_{\gamma 0}$, such that the sum over the models visited is one (i.e. probabilities are obtained by re-normalizing). Notice that this way, MCMC methods act as  searching methods (on the grounds that good models should appear more frequently because they are more probable).

When analyzing the method in the preceding paragraph two observations arise. First, since frequencies are not used, visiting a model more than once is a {\em waste} of time, suggesting that it would be preferable to sample without replacement. The second is that unvisited models in general have very low probability, so we should mainly focus on sampling good models (with high posterior probabilities). These two ideas have inspired the appearance of specific methods to search the model space for good models without repetition. Examples of such proposals are the Bayesian Adaptive Sampling method of \cite{Clyetal10}, the searching method of \cite{BerMol05} in which the Feature Inclusion Stochastic Search (FINCS) of \cite{ScCar09} and \cite{CarSc09} is based. A common recursive idea in these methods is that the exploration of ${\cal M}$ is guided by estimates of inclusion probabilities of single covariates. This potentially leads to biased results because, it does not have to be the case that high inclusion probabilities point to the most probable models. We will see a demonstration of this effect in the examples in Section~\ref{sec.exact} and Section~\ref{sec.larger}.

\section{Inferences in model selection problems}\label{pss}
In a model selection problem, one relevant question is which of the proposed models is the most probable in the light of the data (the highest posterior probability model, HPM). In this situation, the quantity of interest is
$$
\tau=\mbox{HPM}=\argmax_{\gamma\in M} B_{\gamma 0}\, Pr(M_\gamma).
$$
Given a set of visited models ${\cal M}^*$, the obvious and most precise way of estimating the HPM is common to {\em empirical} and {\em re-normalized} methods and is:
$$
\hat\tau=\widehat{\mbox{HPM}}=\argmax_{\gamma\in M^*} B_{\gamma 0}\, Pr(M_\gamma).
$$
Notice that the goodness of the estimation of the HPM only depends upon the ability of the methods to search for good models in very large model spaces.

Nevertheless, very frequently, the quantity of interest $\tau$ which we want to infer is of a different nature. A crucial aspect is that on many occasions this quantity implicitly depends on the normalizing constant. These can be written in terms of the expectation
\begin{equation}\label{total}
\tau(a)=E_{Pr(\cdot\mid y)}(a(M_\gamma))=\sum_{\gamma}\, a(M_\gamma)\, Pr(M_\gamma\mid\n y),
\end{equation}
where $a(M_\gamma)$ is a known function of $M_\gamma$. Clearly, the posterior probability of a single model $M_{\gamma*}$ can be expressed as (\ref{total}) with $a(M_\gamma)=1$ if $M_\gamma=M_{\gamma*}$, and zero otherwise. There are many other examples of such representation of quantities of interest in the model selection problem.

{\small
\paragraph{Example 1: Inclusion probabilities and the median probability model} {\sl For a given explanatory variable $x_l$ the inclusion probability is defined as
$$
q_l=\sum_{\gamma:\,\gamma_l=1}\, Pr(M_\gamma\mid \n y).
$$
These probabilities have interesting theoretical 
properties as shown in \cite{BarBer04} and are useful summaries of the posterior distribution. In particular, they can be helpful when the number of models is very large and the posterior probabilities of single models are so small that are very difficult to interpret.  Apart from their intrinsic interest, inclusion probabilities are the basis of the median probability model in \cite{BarBer04}. This model, hereafter called MPM, is defined as the one with those variables with $q_l>0.5$ and the theory in \cite{BarBer04} suggests that the MPM model has optimal properties and is better for prediction purposes than the HPM  (a surprising fact).
The probability, $q_l$ can be expressed as (\ref{total}) with $a(M_{\gamma})=1$ if $\gamma_l=1$, and 0 otherwise.

Inclusion probabilities are the most popular element in a set of useful summaries for the variable selection problem. For instance, we can be interested in the joint posterior probability of both $x_l$ and $x_{l'}$ and this measure can also bewritten easily in the form of (\ref{total}).
}}

{\small
\paragraph{Example 2: Posterior probability of dimension of the `true' model}  {\sl The probability that the `true' model has exactly $k^*$ explanatory covariates is
$$
d(k^*)=\sum_{\gamma:\,k_\gamma=k^*}\, Pr(M_\gamma\mid \n y).
$$

This corresponds to the expression in (\ref{total}) with $a(M_\gamma)=1$ if $k_\gamma=k^*$, and 0 otherwise.}}

{\small 
\paragraph{Example 3: Model averaging techniques} {\sl Suppose that $\Delta$ is a quantity of interest, then the posterior distribution $Pr(\Delta\mid\n y)$ is ($\ref{total}$) with $a(M_\gamma)=Pr(\Delta\mid\n y,M_\gamma)$. 

What arises is, of course, the methodology called Model Averaging, which is just the Bayesian way of accounting for the uncertainty regarding which the true model is (see eg. \citealt{Hetal99}).

Special mention should be made of the case where $\Delta$ is a future observable $y^{new}$, given certain values of the explanatory covariates $\n x^{new}$. In this case $Pr(y^{new} \mid\n y)$ is the posterior predictive distribution. Notice also that summaries of this distribution are special cases of ($\ref{total}$), like the posterior predictive expectation (with $a(M)=E(y^{new}\mid M_\gamma,\n y)$) or the posterior predictive variance.\\[.5cm]}}

Now suppose ${\cal M}^*=\{\n\gamma^{(1)},\n\gamma^{(2)},\ldots,\n\gamma^{(n)}\}$ have been randomly simulated with replacement from ${\cal M}$ such that on each draw, each model $M_{\gamma}$ has a probability $Pr(M_\gamma\mid \n y)$ of being selected (we think of ${\cal M}^*$ as approximately the sample of models produced with MCMC methods, Section~\ref{sec.slm}). What arises is a probability proportional to size sampling (see \citealt{Lohr99}) where the 'size' of each sampling unit (the models) is $Pr(M_\gamma\mid\n y)$. The usual estimator of $\tau(a)$ in (\ref{total}) under this sampling scheme is
$$
\widehat\tau(a)=\frac{1}{n}\,\sum_{j=1}^n\, \frac{a(M_{\gamma^{(j)}})Pr(M_{\gamma^{(j)}}\mid \n y)}{Pr(M_{\gamma^{(j)}}\mid \n y)}=\frac{1}{n}\,\sum_{j=1}^n\, a(M_{\gamma^{(j)}}),
$$
usually known in the literature as the Hansen-Hurwitz for random sampling with replacement estimator (\citealt{HanHur43}). It can be easily shown that $\widehat\tau(a)$ is an unbiased estimator of $\tau(a)$ (\citealt{Lohr99}).

As a consequence, the Hansen-Hurwitz estimator of the posterior probability of a single model $M^*$ is the frequency of $M^*$ in ${\cal M}^*$. Likewise, Hansen-Hurwitz estimators of the quantities in the previous examples are
$$
\widehat q_l=\frac{1}{n}\sum_{j:\,\gamma^{(j)}_l=1}\, 1,\hspace{1cm}
\widehat d(k^*)=\frac{1}{n}\sum_{j:\,k_{\gamma^{(j)}}=k^*}\, 1,
$$
and
$$
\widehat{Pr}(\Delta\mid\n y)= \frac{1}{n}\sum_{j=1}^n Pr(\Delta\mid\n y,M_{\gamma^{(j)}}).
$$

Of course, these are just the {\em empirical} estimators (as labeled in Section~1) based on the frequency of visits. This correspondence is a key point which provides theoretical support to the arguments introduced in Section~1 and our experience (partially presented in the following sections) regarding the extremely good results of {\em empirical} methods. It now becomes obvious that these estimators are implicitly based on the analytical expression of the Bayes factor through the sampling mechanism used. Moreover, they enjoy the desirable properties of Hansen-Hurwitz estimators, as for example unbiasedness. 

Furthermore, it is quite interesting that these estimators come with a measure of precision, a characteristic that has remained unnoticed in this context until now. This may have interesting applications and important consequences as, for instance, knowing when $n$ gives enough precision in the estimation of the quantity of interest. If the draws on ${\cal M}^*$ are independently obtained, the variance of $\hat\tau(a)$ is (see eg. \citealt{Lohr99})
$$
V(\hat\tau(a))=\frac{1}{n}\sum_{\gamma\in{\cal M}}\, Pr(M_\gamma\mid\n y)\big(a(M_\gamma)- \tau(a)\big)^2.
$$
In the case that the quantity of interest is a probability $p$ (e.g. probability of a single model or an inclusion probability), $V(\hat\tau(a))=n^{-1}p(1-p)$ which is, of course, bounded above by $1/(4n)$ (this bound being a reasonable measure of the variability for probabilities not very close to zero).
This provides an accurate idea of the precision achieved with the procedure and can be used (depending on the magnitude of the probability being estimated), for example, to decide the 
number of draws needed. 

Also useful is that an unbiased estimator of the variance of $\hat\tau(a)$ is
\begin{equation}\label{var.tau}
\hat{V}(\hat\tau(a))=\frac{1}{n(n-1)}\, \sum_{j=1}^n\, \big(a(M_{\gamma^{(j)}})-\hat\tau(a)\big)^2.
\end{equation}

At this point it could argued that these results are of limited importance in practice since in MCMC sampling schemes we are not exactly sampling from the posterior and draws are dependent. Strictly speaking this is true, although our experience (partially shown in the following sections) is that these properties (the unbiasedness of $\hat\tau(a)$ and the expression for $\hat{V}(\hat\tau(a)$) hold quite accurately in practice. On the other hand, notably, these are basic assumptions that underlie {\em any} analysis solved with MCMC methods, the literature containing plenty of techniques  for improving the results of MCMC methods in this sense. Among them, probably the most popular and simple to implement yet very effective are {\em thinning} (to systematically keep one simulation out of several) and {\em burning} (to reject some of the first simulations).

The estimator of $\tau(a)$ within the {\em re-normalized} approach is
$$
\sum_{\gamma\in M^*}\, a(M_\gamma)\, \widehat{Pr}(M_\gamma\mid\n y)
$$
where the posterior probabilities of single models are obtained by re-normalizing the Bayes factors, that is
$$
\widehat{Pr}(M_\gamma\mid\n y)= B_{\gamma 0}/ \sum_{\gamma\in M^*}\,B_{\gamma 0}.
$$
The properties of these estimators are in general difficult to derive. The bias of such estimators for the posterior probability for single models has been the subject of a recent study in \cite{ClyGho10}.

\section{Example I: a large problem with an exact solution}\label{sec.exact}
Mainly for comparative purposes but also to report the exact results on a moderately large problem (something that has not been done before to the best of our knowledge), here we present the exact solution for a problem with $p=35$ covariates, and hence with $34,359,738,368\approx 3\cdot10^{10}$ different models. Having the exact results of a large problem seems to us the most informative and clarifying way of comparing the performance of searching methods. 

\begin{table}[t!]
\begin{center}
{\small\scalebox{0.75}{
\begin{tabular}{cl}
Variable & Description \\
\hline
$y$ & Response = Daily maximum 1-hour-average ozone reading (ppm) at Upland, CA\\
$x_1$ & Month: 1 = January, . . . , 12 = December\\
$x_2$ & Day of month\\
$x_3$ & Day of week: 1 = Monday, . . . , 7 = Sunday\\
$x_4$ & 500-millibar pressure height (m) measured at Vandenberg AFB\\
$x_5$ & Wind speed (mph) at Los Angeles International Airport (LAX)\\
$x_6$ & Humidity (\%) at LAX\\
$x_7$ & Temperature (Fahrenheit degrees) measured at Sandburg, CA\\
$x_8$ & Inversion base height (feet) at LAX\\
$x_9$ & Pressure gradient (mm Hg) from LAX to Daggett, CA\\
$x_{10}$ & Visibility (miles) measured at LAX\\
\hline
\end{tabular}
}
\caption{{\small Description of variables used in Example I and Example II}}
\label{descrip}}
\end{center}
\end{table}

It is generally understood that problems with $p$ larger than 25-30 are intractable, \citep{Clyetal10}, and we are considering a step beyond this limiting size. The results we obtained were derived using a cloud with 150 processors and took approximately 20 hours to run. The code was written in C, with the gsl library \citep{Galetal09}. The source code is available upon request.

The data we analyzed were previously used by \cite{CasMor06} and \cite{BerMol05} and concern $N=178$ measures of ozone concentration in the atmosphere. Details on the data can be found in \cite{CasMor06}. Of the 10 main effects originally considered, we only make use of those with an atmospheric meaning, as was done by \cite{liang08}. Then we have 7 main effects which, jointly with the quadratic terms and second order interactions, produce the above mentioned $p=35$ possible regressors. For comparative purposes, we keep the original notation of the variables defined in Table~\ref{descrip}. We call this dataset Ozone35, for which we now present the exact results. We use the $g$-prior with $g=N$ and a constant prior for the prior probabilities of models.

The sum of all Bayes factors (the proportionality constant) is
$$
\sum_{\gamma}\, B_{\gamma0}(g)=1.13\, 10^{50}.
$$
The highest probability model, HPM, has covariates $\{1,x_{10},x_4x_6,x_6x_8,x_7^2,x_7x_{10}\}$ and has a posterior probability of $0.0009$, with a Bayes factor (in its favor and against $M_0$) of $1.02\, 10^{47}$. The first 1000 most probable models accumulated a total probability of $0.07$ and a sum of Bayes factors (expressed in decimal logarithm) of 48.92 (this value is used later). Inclusion probabilities of each variable are in Table~\ref{incprob}. Hence, the median inclusion probability model, MPM, is
$\{1,x_6^2,x_6x_7,x_6x_8,x_7x_{10}\}$ which has a posterior probability which is twenty three times lower than the probability of the highest posterior probability. Moreover, there are 851 models which are more probable than the median inclusion probability model.

We then run the following methods ten times, each with $n=10000$ iterations.
\begin{itemize}
\item[{\tt Freq}] Gibbs sampling with algorithm in Appendix~\ref{ApGS} the with initial model $M^{(0)}=M_0$ (we did not observe differences starting with the full model or with a randomly chosen model). For a fair comparison among the methods compared we did not exclude any model sampled and did not use any burning period.

\item[{\tt BAS}] Bayesian adaptive sampling of \cite{Clyetal10} through the corresponding R-package ${\tt BAS}$. As recommended (personal communication), we used ${\tt method="MCMC+BAS"}$, which uses an MCMC method to initialize the search (this is a clear improvement over other options like ${\tt eplogp}$, which uses a rough approximation of inclusion probabilities with p-values to initialize the search). We tuned the parameter ${\tt update=500}$ so that sampling probabilities were updated every 500 iterations.

\item[{\tt SSBM}] The Stochastic Search in \cite{BerMol05}. This method was originally proposed for a particular prior but the searching algorithm can be easily adapted to accommodate the $g$-prior.

\end{itemize}

The estimates computed in {\tt Freq} are proportional to size (see Section~\ref{pss}) and hence based on frequency of visits, and in {\tt BAS} and {\tt SSBM} estimators are based on the renormalization of the Bayes factors. Hence, {\tt Freq} is a particular method within the {\em empirical} approach, while {\tt BAS} and {\tt SSBM} are methods of the {\em re-normalized} approach (using the labels introduced in Section~\ref{sec.est}). The results are summarized in Table~\ref{incprob}.

For the first run of each method,  in Table~\ref{incprob} we present estimates of the inclusion probabilities, the MPM and the HPM. With this same run we estimated the standard deviation of the estimators of the inclusion probabilities with {\tt Freq} using (\ref{var.tau}). In addition, with the ten runs we computed the observed standard deviation as this provides a measure of variability in {\tt BAS} and {\tt SSBM} (for which an expression like the one in \ref{var.tau} does not exist).

The main conclusions that we have extracted from the former simulations can be summarized as follows:

\paragraph{Regarding the MPM and inclusion probabilities}
Of the ten experiments conducted, {\tt Freq} correctly identified the MPM  ten times while with {\tt BAS} and {\tt SSBM} the estimated MPM and the real MPM did not coincide in any of the ten runs. Also, {\tt Freq} provides very accurate estimations of the (exact) inclusion probabilities with a small variability. This confirms the high efficiency of such estimators. The observed variability with {\tt BAS} and {\tt SSBM} is large, so in general we expect large differences in repetitions of these methods in a similar manner to that observed in this experiment.

One great advantage of {\em empirical} methods over {\em re-normalized} is that the first come with a measure of precision in the estimates (\ref{var.tau}). This measure can be legitimately criticized since it is just an approximation due to the dependency between the simulations. Nevertheless, in this experiment these estimates and the observed standard deviation are quite close to each other, suggesting that  (\ref{var.tau}) is quite a reasonable estimator.

The most worrisome aspect observed of {\tt BAS} and {\tt SSBM} is that they are clearly biased: for certain covariates we have to move from the point estimation more than 10 times the standard deviation to cover the exact value of the inclusion probabilities (see eg. $x_6x_8$ in {\tt BAS} and $x_5x_{10}$ in {\tt SSBM}). The nature and origin of this bias has an easy interpretation after a careful reading of the table. In {\tt BAS}, six inclusion probabilities are overestimated, of which five are in the estimated HPM; the rest are underestimated. A similar pattern is found in {\tt SSBM}. This means that inclusion probabilities within these methods are very influenced by the estimated HPM, leading to a bias in the direction of the HPM model.
This effect is, in our opinion, the manifestation of a search in the model space for good models guided by the inclusion probabilities.

\paragraph{Regarding the HPM and probability mass discovered}
One interesting question is which method is visiting better models. {\tt BAS} and {\tt SSBM} are, in some sense, specifically designed with this aim while this characteristic is presumed in MCMC methods (since more probable models should be visited just because they are more probable). In our experiment, {\tt BAS} correctly identified the HPM  nine times while {\tt SSBM} did it five times. The exact HPM was among the visited models in {\tt Freq} in the ten runs, showing that {\tt Freq} is visiting good models. 

Finally, we calculated the mean and standard deviation (over the ten runs) of the sum of the Bayes factors of the 1000 (in decimal logarithm) most probable different models explored (to be compared with the exact value given above of 48.92). The results were 48.77(0.01), 48.64(0.05) and 48.50(0.20), for {\tt Freq}, {\tt BAS} and {\tt SSBM} respectively. In this respect, the three methods analyzed behave similarly, although {\tt Freq} gives more stable answers.


\begin{table}[t!]
\begin{center}
{\small\scalebox{0.75}{
\begin{tabular}{c|c|cccccccccc}
&Method & $1$              & $x_4$ & $x_5$ & $x_6$ & $x_7$ & $x_8$ & $x_9$ & $x_{10}$ & $x_4^2$ \\
\hline
$q_l$ & exact & 1$*\dag$         & 0.164 & 0.096 & 0.297 & 0.195 & 0.200 &  0.291 & 0.368* & 0.164\\
$\hat q_l$ & {\tt Freq} & 1$*\dag$    & 0.157 & 0.099 & 0.300 & 0.191 & 0.200 & 0.292 & 0.368* & 0.162\\
$[\hat V(\hat q_l)]^{1/2}$ & {\tt Freq} & (0)     & (0.004) & (0.003) & (0.005) & (0.004) & (0.004) & (0.005) & (0.005) & (0.004)\\
$S(\hat q_l)$ & {\tt Freq} & (0)     & (0.005)& (0.002)& (0.007)& (0.008)& (0.007)& (0.004)& (0.005)& (0.004)\\
$\hat q_l$ & {\tt BAS}  &  1$*\dag$   &  0.022 &  0.01 &  0.231 &  0.032 &  0.025 &  0.092 &  0.508$*\dag$ &  0.023\\
$S(\hat q_l)$ & {\tt BAS}  & (0)     & (0.007) & (0.003) & (0.046) & (0.017) & (0.018) & (0.027) & (0.078) & (0.006)\\
$\hat q_l$ & {\tt SSBM} &  1$*\dag$   &  0.105 &  0.03 &  0.04 &  0.053 &  0.073 &  0.297 &  0.131 &  0.125\\
$S(\hat q_l)$ & {\tt SSBM}  & (0)    & (0.034) & (0.006) & (0.154) & (0.021) & (0.046) & (0.086) & (0.277) & (0.037)\\[.1cm]
& & $x_4x_5$ & $x_4x_6$ & $x_4x_7$ & $x_4x_8$ & $x_4x_9$ & $x_4x_{10}$ & $x_5^2$ & $x_5x_6$ & $x_5x_7$ \\
\hline
$q_l$ & exact & 0.095 & 0.325* & 0.252 & 0.208 & 0.301  & 0.361 &  0.124 &  0.107 & 0.094\\
$\hat q_l$ & {\tt Freq} & 0.094 & 0.320* & 0.244 & 0.210 & 0.303 & 0.360 & 0.127 & 0.104 & 0.095\\
$[\hat V(\hat q_l)]^{1/2}$ & {\tt Freq} & (0.003)& (0.005)& (0.004)& (0.004)& (0.005)& (0.005)& (0.003)& (0.003)& (0.003)\\
$S(\hat q_l)$ & {\tt Freq} & (0.002)& (0.01)& (0.006)& (0.005)& (0.008)& (0.006)& (0.002)& (0.003)& (0.003)\\
$\hat q_l$ & {\tt BAS}  &  0.019 &  0.373* &  0.164 &  0.061 &  0.078 &  0.416 &  0.019 &  0.013 &  0.012\\
$S(\hat q_l)$ & {\tt BAS}  & (0.003) & (0.09) & (0.043) & (0.024) & (0.049) & (0.061) & (0.005) & (0.004) & (0.003)\\
$\hat q_l$ & {\tt SSBM} &  0.037 &  0.03 &  0.082 &  0.092 &  0.348* &  0.132 &  0.047 &  0.049 &  0.035\\
$S(\hat q_l)$ & {\tt SSBM}  & (0.007) & (0.295) & (0.285) & (0.16) & (0.098) & (0.33) & (0.008) & (0.011) & (0.008)\\[.1cm]
& &  $x_5x_8$ & $x_5x_9$ & $x_5x_{10}$ & $x_6^2$ & $x_6x_7$ & $x_6x_8$ & $x_6x_9$ & $x_6x_{10}$ & $x_7^2$ \\
\hline
$q_l$ & exact &  0.098&  0.088 & 0.124 & 0.532$\dag$ &  0.636$\dag$ &  0.560$*\dag$ & 0.126 &  0.115 &  0.450*\\
$\hat q_l$ & {\tt Freq} & 0.098 & 0.087 & 0.124 & 0.524$\dag$ & 0.634$\dag$ & 0.564$*\dag$ & 0.127 & 0.113 & 0.465*\\
$[\hat V(\hat q_l)]^{1/2}$ & {\tt Freq}& (0.003)& (0.003)& (0.003)& (0.005)& (0.005)& (0.005)& (0.003)& (0.003)& (0.005)\\
$S(\hat q_l)$ & {\tt Freq}  & (0.002)& (0.003)& (0.004)& (0.008)& (0.012)& (0.007)& (0.003)& (0.001)& (0.009)\\
$\hat q_l$ & {\tt BAS}  & 0.009 &  0.014 &  0.014 &  0.282 &  0.493 &  0.929$*\dag$ &  0.025 &  0.019 &  0.793$*\dag$\\
$S(\hat q_l)$ & {\tt BAS}  & (0.003) & (0.003) & (0.004) & (0.078) & (0.117) & (0.034) & (0.004) & (0.007) & (0.066)\\
$\hat q_l$ & {\tt SSBM} &  0.027 &  0.017 &  0.023 &  0.98$*\dag$ &  1$*\dag$ &  0.077 &  0.078 &  0.031 &  0.112\\
$S(\hat q_l)$ & {\tt SSBM} & (0.005) & (0.01) & (0.009) & (0.301) & (0.37) & (0.339) & (0.017) & (0.043) & (0.342)\\[.1cm]
& &  $x_7x_8$ & $x_7x_9$ & $x_7x_{10}$ & $x_8^2$ & $x_8x_9$ & $x_8x_{10}$ & $x_9^2$ & $x_9x_{10}$ & $x_{10}^2$\\
\hline
$q_l$ & exact & 0.349 & 0.431 &  0.743$*\dag$ & 0.142 & 0.263 & 0.236 & 0.434 & 0.103 & 0.117\\
$\hat q_l$ & {\tt Freq} & 0.346 & 0.430 & 0.756$*\dag$ & 0.140 & 0.264 & 0.231 & 0.440 & 0.103 & 0.116\\
$[\hat V(\hat q_l)]^{1/2}$ & {\tt Freq}& (0.005)& (0.005)& (0.004)& (0.003)& (0.004)& (0.004)& (0.005)& (0.003)& (0.003)\\
$S(\hat q_l)$ & {\tt Freq} & (0.008)& (0.01)& (0.006)& (0.004)& (0.008)& (0.004)& (0.005)& (0.003)& (0.002)\\
$\hat q_l$ & {\tt BAS}  & 0.091 &  0.124 &  0.965$*\dag$ &  0.017 &  0.045 &  0.127 &  0.393 &  0.022 &  0.018\\
$S(\hat q_l)$ & {\tt BAS}  & (0.047) & (0.073) & (0.026) & (0.013) & (0.032) & (0.028) & (0.032) & (0.005) & (0.007)\\
$\hat q_l$ & {\tt SSBM} &  0.975$*\dag$ &  0.663$*\dag$ &  0.879$*\dag$ &  0.026 &  0.57$*\dag$ &  0.597$*\dag$ &  0.244 &  0.059 &  0.064\\
$S(\hat q_l)$ & {\tt SSBM} & (0.425) & (0.193) & (0.209) & (0.01) & (0.164) & (0.178) & (0.084) & (0.015) & (0.015)\\
\hline
\end{tabular}
}
\caption{{\small Inclusion probabilities (exact $q_l$ and estimates $\hat q_l$ in one run of 10000 iterations) for the Ozone35 data set. Also, $\hat V(\hat q_l)$ is the estimated variance (\ref{var.tau}) using this same run and $S(\hat q_l)$  is the deviation of the estimators observed in ten independent identical runs. Symbols ($\dag$) for those variables in the estimated MPM and asterisks ($*$) for those variables in the estimated HPM.}}
\label{incprob}}
\end{center}
\end{table}

\section{Example II: A much larger problem}\label{sec.larger}
We now consider the full Ozone dataset with the 10 main effects, the quadratic terms and the second order interactions. The same problem has been considered by \cite{BerMol05} and, as before, we keep the same notation for the covariates as there. This problem has $p=65$ and hence $2^{65}\approx 3.7\, 10^{19}$ models in ${\cal M}$. In what follows we call this dataset Ozone65.  The size of ${\cal M}$ precludes having the exact answer to the problem. To have an approximate idea of the unfeasibility, notice that with the C code that we used for the Ozone35 it would take more than 350 years to compute the answer using a cloud with $10^6$ processors. We cannot wait that long.

For this dataset we repeated the comparison in Section~\ref{sec.exact} and performed 10 different runs, now each with $n=100,000$ iterations, of {\tt Freq}, {\tt BAS} and {\tt SSBM}. In Table~\ref{oz65incprob} we present the statistics of all the variables included in the estimated HPM and MPM in any of the runs of the methods being compared. In essence, these results are in clear agreement with our findings in Ozone35, and confirm the conclusions drawn there.

\paragraph{Regarding the MPM and inclusion probabilities}
It is unknown which model is the MPM (and it will probably never be known) but results with {\tt Freq} provide a very reasonable and consistent picture of the solution. In {\tt Freq}, except for one run, there is unanimity in the estimation of the MPM. Furthermore, and quite appealing is that we can give an explanation of the disagreement in terms of errors in the estimation. The discordant run differs from all the others in that it includes $x_1x_4$. This variable has an estimated inclusion probability of $0.497$ with an estimated error of $0.002$. 

In {\tt BAS} and {\tt SSBM} results vary considerably over the different runs. In {\tt BAS} ({\tt SSBM}) 8(10) different models were estimated as the MPM and hence, at least 8(9) times this method has incorrectly identified the true MPM. More worrisome is that these bad results do not seem to be due to variability. We can find a more likely explanation in  Table~\ref{oz65incprob} where we can clearly see that, in many occasions, the estimated MPM mimics the estimated HPM. For instance, $x_6$ and $x_4x_6$ (which are in the estimated HPM) are always in the MPM estimated by {\tt BAS}. This also seems to be the case for $x_4x_7$ and $x_6x_7$ in {\tt SSBM}. We interpret these results as a manifestation of the bias produced with methods conceived to look for good models.

\paragraph{Regarding the HPM and probability mass discovered}
In this aspect, the three methods behave quite similarly, perhaps {\tt BAS} and {\tt Freq} perhaps performing slightly better than {\tt SSBM}.
The best model found in the whole experiment, the ten runs of the three different methods, had a Bayes factor (in its favor and against the null) of 50.87 (expressed in decimal logarithm). This model, identified in Table~\ref{oz65incprob} with asterisks, was visited in four of the ten runs by {\tt BAS}, in three runs by {\tt Freq} and in one run by {\tt SSBM}. The means (standard deviations) over the ten runs of the sum of the Bayes factors of the 1000 (in decimal logarithm) most probable different models explored were 52.77(0.02) in {\tt Freq}, 52.78(0.15) in {\tt BAS} and 52.78(0.16) in {\tt SSBM}. These results confirm the popular hypothesis that good models also show up when sampling from the posterior distribution.

\begin{table}\begin{center}
{\small\scalebox{0.75}{
\begin{tabular}{c|ccc|ccc|c} 
 & \multicolumn{3}{c}{HPD} & \multicolumn{3}{c}{MPM}& $\hat{q}_l\,(\hat{V}(\hat{q}_l)^{1/2})$\\
\hline
Method & Freq & BAS & SSBM & Freq & BAS & SSBM&\\
\hline
$x_1$*       & 7  & 7  & 8     & 10 & 8    & 8 & 0.575(0.002)\\
$x_6$*       & 8  & 9  & 4     & -   & 10   & 2  & 0.427(0.002)\\
$x_7$       &  -    & -      &  3   &  -  & -     & 4  & 0.290(0.001)    \\
$x_8$       & 1    & -     & 1    & -    & -    & 1    & 0.297(0.001)   \\
$x_{10}$*     & 3  & 5  & 4      & -    & 2    & 2   & 0.292(0.001)    \\
$x_1 x_1$* & 10& 10& 10    & 10 & 10  & 10 & 1.000($<$0.001)\\ 
$x_1x_4$  & 3    & 3    & 2     & 1   & 3   & 2     & 0.497(0.002)\\
$x_2x_8$*  & 9    & 8  & 4   & -    & -    & -   & 0.184(0.001) \\
$x_4x_4$  &  -    & -      & 1   &  -    & -   & 1     & 0.266(0.001)   \\
$x_4x_6$*  & 8  & 9  & 1     & -    & 10  & 2   & 0.418(0.002)\\
$x_4x_7$  & 2    & 1    & 5     & -    & -    & 5    & 0.337(0.001)  \\
$x_4x_8$  & -    & 2    & 3      & -    & 1   & 3    & 0.303(0.001)  \\
$x_4x_{10}$ & 6    & 3   & 4     & -    & 2   & 4   & 0.309(0.001) \\
$x_5x_5$*  &10    & 9  & 9  & -     & -   & -   & 0.261(0.001) \\
$x_5x_7$  &   -    & 1    &  -    &  -     & -   &  -     & 0.156(0.001) \\
$x_6x_6$  &  -    &  -     &   -   &  -    &  -  &  2   & 0.218(0.001)  \\
$x_6x_7$  & 1    & 2    & 7     & 10  & 3  & 7     & 0.614(0.002)\\
$x_6x_8$  &  -    & -      &   -   &  -    & -   & 6   & 0.328(0.001)   \\
$x_6x_{10}$& 1    & 1    & 2     & -     & -   & 2     & 0.237(0.001)  \\
$x_7x_7$*  & 7  & 7  & 2     & -     & 6  & 2    & 0.372(0.002)\\
$x_7x_8$  & 1    & 2    & 2     & -     & 2  & 2     & 0.479(0.002)\\
$x_7x_{10}$* & 9  & 9  & 8     & 10   & 9  & 8    & 0.623(0.002)\\
$x_9x_9$*  & 10& 10&  10   & 10  & 10 &  10 & 0.966(0.001) \\
\hline
& \multicolumn{6}{c}{Number of different variables}\\
& 17 & 18 & 20 & 6 & 13 & 20
\end{tabular}
}
\caption{{\small For the Ozone65 dataset, the number of times each covariate is included in the estimated HPM and the estimated MPM in ten independent runs of $n=100000$ iterations of {\tt Freq}, {\tt BAS}, and {\tt SSBM} methods. Asterisks identify the best model encountered in the full experiment. Also, $\hat{q}_l$'s are the estimation of the inclusion probabilities from the first run of {\tt Freq} and $\hat V(\hat q_l)$ is their estimated variance (\ref{var.tau}) using this same run.}}
\label{oz65incprob}}\end{center}
\end{table}

\section{Summary and main conclusions}\label{sec.Ext}
In the context of Bayesian model selection with very large model spaces ${\cal M}$, quantities of interest $\tau$ in the problem have to be estimated since their exact value is, in practice, unknown. This is mainly because the underlying normalizing constant, whose determination would imply the computation of Bayes factors for all the models, is virtually unknown. In this situation, such estimates have to be constructed from a sample of models ${\cal M}^*$ of ${\cal M}$ and can be derived either by using the empirical distribution
or through renormalization of the Bayes factors. Within the first approach, ${\cal M}^*$ has to be a sample from the posterior distribution and is usually obtained with MCMC sampling. In the second approach, ${\cal M}^*$ does not necessarily have to be obtained with probabilistic-based mechanisms and the emphasis is normally placed on sampling good models. We labeled each of these approaches {\em empirical}  and {\em re-normalized}. We have shown that {\em empirical} estimates are in general, under the common assumptions in MCMC sampling, unbiased. Also, the uncertainty regarding the estimations can be easily derived, making the {\em empirical} approach very appealing.

We have compared several methods in a moderate to large problem with $p=35$ covariates for which we have derived the exact answers, and on a larger problem with $p=65$ with an unknown solution. With respect to sampling good models and in particular the highest posterior probability model, (a problem for which the normalizing constant is not needed), {\em empirical} and {\em re-normalized} methods behave quite similarly. Nevertheless, in the estimation of other important parameters like the inclusion probabilities, {\em re-normalized} methods can be strongly biased.  

\section*{Acknowledgements} We want to thank Rafael Espinosa at the supercomputing center in the Institute for Research in Information Technology at 
the Universidad de Castilla-La Mancha for providing us with technical support. This work has been partially funded by a project granted by the Spanish Ministry of Science and Education coded MTM2010-19528.

\appendix
\section{Gibbs sampling algorithm for hierarchical $g$-priors}\label{ApGS}

Once the parameters $\n\beta_\gamma,\sigma$ have been analytically integrated out (see \ref{gBF}), the only unknown parameters in the problem are $g$ and the components in $\n\gamma$. Those have full conditional distributions:
\begin{equation}\label{FullCH}
\gamma_i\mid \gamma_1,\ldots,\gamma_{i-1},\gamma_{i+1},\ldots,\gamma_{p},g,\n y\sim \mbox{Bernoulli}(p_i),
\end{equation}
where
$$
p_i=\frac{B_{a 0}(g)\pi(g\mid\n\gamma=\n a)Pr(M_{a})}{B_{a 0}(g)\pi(g\mid\n\gamma=\n a)Pr(M_{a})+B_{b0}(g)\pi(g\mid\n\gamma=\n b)Pr(M_{b})},
$$
where
$$ 
\n a=(\gamma_1,\ldots,\gamma_{i-1},1,\gamma_{i+1},\ldots,\gamma_{p}),
$$ 
and
$$ 
\n b=(\gamma_1,\ldots,\gamma_{i-1},0,\gamma_{i+1},\ldots,\gamma_{p}).
$$ 
The full conditional for $g$ is
$$
f(g\mid\n\gamma, \n y)\propto B_{\gamma 0}(g)\, \pi(g\mid\n\gamma),
$$
which can easily be sampled via Metropolis-Hastings with an obvious proposal: $g^*\sim\pi(g\mid\n\gamma)$.
The case for $g$-priors and any other prior that leads to a closed expression is a particular case of the above with $\pi(g\mid\n\gamma)=1$ and the step for $g$ is not used.

\bibliography{$HOME/Mywork/bibliografia/MSComputation,$HOME/Mywork/bibliografia/Anabel}

\bibliographystyle{plainnat}

\end{document}